\documentclass[twocolumn]{IEEEtran}
 \usepackage{url,amsmath,graphicx}
\usepackage{amsmath,amssymb,amscd,latexsym,dsfont,amsthm}
\usepackage{cite}
\usepackage{amsmath,amssymb,amsfonts}
\usepackage{algorithmic}

\usepackage{graphicx}
\usepackage{textcomp}
\usepackage{xcolor}

\usepackage{hyperref} 
\usepackage{graphicx} 
\usepackage[numbers]{natbib}
\usepackage{xcolor}
\usepackage{float}
\usepackage{stfloats}
\usepackage{amsmath}
\usepackage{subcaption}

\usepackage[ruled,vlined]{algorithm2e}
 \usepackage{algorithmic}
  \usepackage{stfloats}
  \usepackage{eqparbox}
   \usepackage{array}
   \usepackage{fixltx2e}
   \usepackage{stfloats}
    \usepackage{cite}
    \usepackage{url}
 \usepackage{cite}
 \usepackage{graphicx}
 \usepackage{psfrag}
 \usepackage{url}
 \usepackage{stfloats}
 \usepackage{array}
 \usepackage{fancyhdr}
 \usepackage{psfrag}
 \usepackage{graphicx}
 \usepackage{lipsum}
 \usepackage{amssymb}
 \usepackage{color}
 \usepackage{subcaption}
 \usepackage{caption}
\usepackage[a4paper,top=0.75in,bottom=1in,left=0.625in,right=0.625in]{geometry}
\begin{document}
\pagenumbering{gobble}

\title{Space Shift Keying-Enabled ISAC for Efficient Debris Detection and Communication in LEO Satellite Networks\\
}
\author{
\IEEEauthorblockN{Gédéon Ghislain Nkwewo Ngoufo, Khaled Humadi, Elham Baladi, Gunes Karabulut Kurt}\\
\IEEEauthorblockA{{\textit{Poly-Grames Research Center, Department of Electrical Engineering} }\\
\text{Polytechnique Montréal, Montréal, QC, Canada}\\
\text{ E-Mails : \{nkwewo-ngoufo.gedeon-ghislain, khaled.humadi, elham.baladi, gunes.kurt\}@polymtl.ca}\\}
\vspace{-1cm}
}
\maketitle
\begin{abstract}
The proliferation of space debris in low Earth orbit (LEO) presents critical challenges for orbital safety, particularly for satellite constellations.
Integrated sensing and communication (ISAC) systems provide a promising dual-function solution by enabling both environmental sensing and data communication. This study explores the use of space shift keying (SSK) modulation within ISAC frameworks, evaluating its performance when combined with sinusoidal and  chirp radar waveforms. SSK is particularly attractive due to its low hardware complexity and robust communication performance. Our results demonstrate that both  waveforms  achieve comparable bit error rate (BER) performance under SSK, validating its effectiveness for ISAC applications. However, waveform selection significantly affects sensing capability: while the sinusoidal waveform supports simpler implementation, its high ambiguity limits range detection. In contrast, the chirp waveform enables range estimation and provides a modest improvement in velocity detection accuracy. These findings highlight the strength of SSK as a modulation scheme for ISAC and emphasize the importance of selecting appropriate waveforms to optimize sensing accuracy without compromising communication performance. This insight supports the design of efficient and scalable ISAC systems for space applications, particularly in the context of orbital debris monitoring.

\end{abstract}
\vspace{-0.2cm}
\section{INTRODUCTION}
The rapid expansion of space activities, particularly the deployment of numerous small satellites in low Earth
orbit (LEO), has led to a significant increase in space debris. These debris, varying in size and origin, pose a critical threat to operational satellites and space missions. As the density of space debris increases, the likelihood of collisions and subsequent cascading effects, known as the Kessler syndrome \cite{1}, becomes more probable. This scenario underscores the urgent need for robust and efficient debris detection systems that can ensure the sustainability and safety of space operations. 

In literature, Integrated sensing and
communication (ISAC) has mainly been explored for LEO satellites for wireless networks, targeting base stations for next-generation networks like 6G. Preliminary research indicates that ISAC can effectively manage interference and enable strong communication and sensing performance with ground applications, which is promising for LEO satellites \cite{yin2024integrated}. Different waveforms have also been explored for communication-centric, sensing-centric and joint applications. The joint application has shown the possibility to achieve great performance trade-off between the communication and sensing functionalities \cite{3}. The full-duplex waveform design has also been explored in previous work \cite{4}, as a great solution to increase the communication rate and target detection but presents greater vulnerability to interference and self-interference in addition on being more complex systems. However, the application of ISAC for LEO satellite constellations needs further investigation for space debris detection scenarios.
Notably, \cite{boudjelal2025single} demonstrated that single carrier  systems are highly energy efficient for integrated radar and communication, making them suitable for long-range sensing applications. The author in \cite{cheng2022integrated} examine  the flexibility of ISAC to be adapted for high-mobility environments, such as vehicular communications.

Index Modulation (IM) has further enhanced the capabilities of ISAC systems by encoding additional information into indices of subcarriers, antennas, or spatial paths. Studies like \cite{hawkins2024ofdm} highlight how IM-Orthogonal Frequency Division Multiplexing (IM-OFDM) systems optimize spectral efficiency while maintaining robust  sensing performance. Similarly, spatial path index modulation (SPIM) techniques \cite{elbir2024spim} have proven effective for mmWave/THz frequencies, enabling precise debris tracking and high data rates in LEO constellations. 



However, many existing studies overlook the unique constraints of LEO constellations, including high mobility, Doppler effects, and multi-orbit dynamics. Addressing these gaps, recent works \cite{liao2024pulse} propose innovative waveform designs and optimization strategies tailored to LEO-specific requirements, balancing spectral efficiency, hardware complexity, and system reliability. 
Despite progress, challenges such as decoding complexity and synchronization persist. This highlights the importance of implementing flexible systems that can adapt by selecting the most suitable waveform for the specific field of application, ensuring optimal communication and detection performance.

The main contributions of this work include the introduction of an ISAC system with space shift keying (SSK) IM for LEO satellites, addressing both orbital debris detection and reliable communication through a low-complexity, spatially encoded modulation scheme. It is important to note that implementing IM-OFDM, as presented in \cite{hawkins2024ofdm}, within the LEO satellite context poses significant challenges. These include increased receiver complexity and increased sensitivity to Doppler spread, which arises because of the high mobility inherent in LEO satellite orbits. IM-OFDM systems require precise frequency synchronization and careful management of the cyclic prefix, both of which become more difficult under severe Doppler shifts. In contrast, the proposed SSK-based approach offers a notably simpler hardware architecture and enhanced robustness to mobility-induced impairments, rendering it more suitable for constrained LEO satellite platforms.
This study analyzes BER performance in SSK-enabled ISAC systems, demonstrating that chirped and sinusoidal waveforms achieve similar BER regardless of the number of transmit antennas, offering flexibility in waveform selection based on sensing accuracy or simplicity. Monte Carlo simulations validate system performance, confirming that while both waveforms yield identical BER, the chirped waveform  outperforms the sinusoidal sensing capabilities.
\section{SYSTEM DESCRIPTION}

\begin{figure}[t]
    \centering   \includegraphics[width=0.5\textwidth]{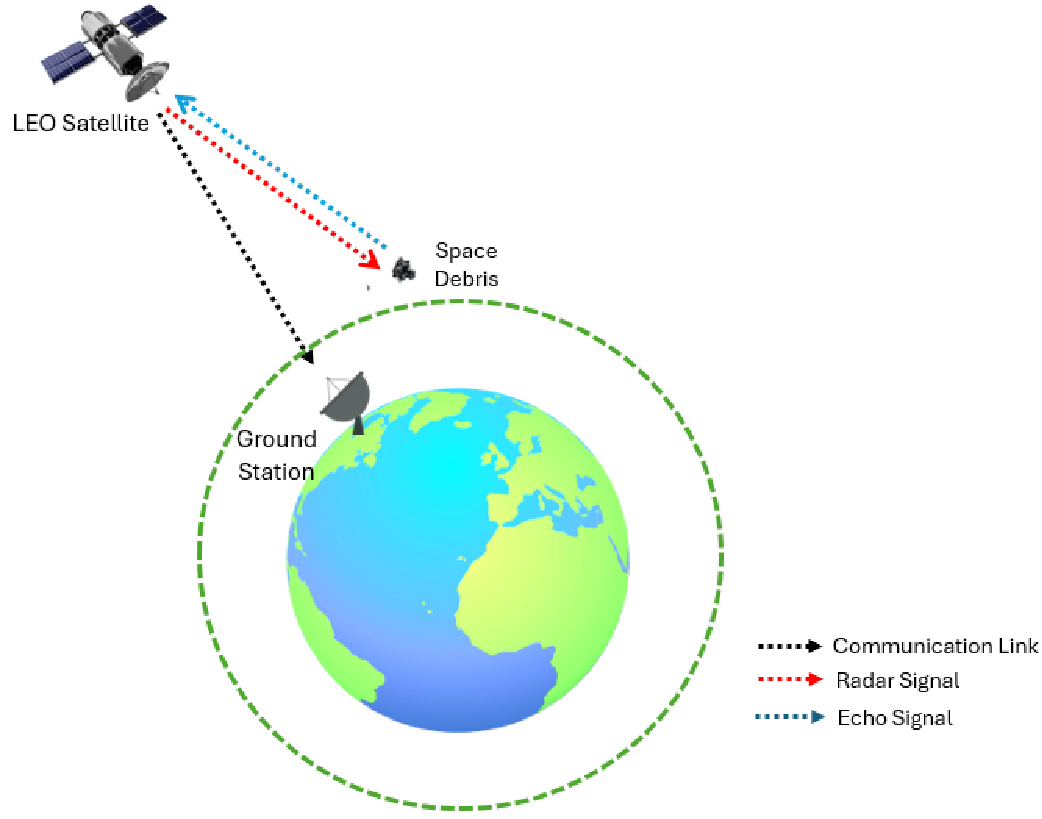}
    \caption{Proposed ISAC system model.
    \vspace{-4mm}
}
    \label{model}
\end{figure}

Linear frequency modulated (LFM) radar signals are critical in ISAC systems designed for LEO satellites. While pulse radars are effective for range measurement, their inability to transmit and receive simultaneously creates blind spots \cite{SalazarAquino2021}, rendering them unsuitable for detecting nearby debris. In contrast, Frequency Modulated Continuous Wave (FMCW) radars transmit continuous chirp signals, eliminating blind spots but lacking the waveform diversity required for robust communication.
Our proposed ISAC system bridges these limitations by adopting an FMCW configuration with adjusted pulse durations to ensure echoes are received within the pulse period. Additionally, it incorporates standard sinusoidal waveforms, as well as SSK modulation and demodulation to enhance communication. This dual-functional framework integrates radar sensing and communication capabilities, offering a cost-efficient and scalable solution for LEO satellite networks. The system's overall architecture is depicted in Fig.~\ref{model}. In the following, we provide a detailed description of the system model components. 
\vspace{-0.5cm}
\subsection{Transmitter}
\begin{figure*}[!t]
    \centering
    \includegraphics[width=1\textwidth,height=0.3\textheight]{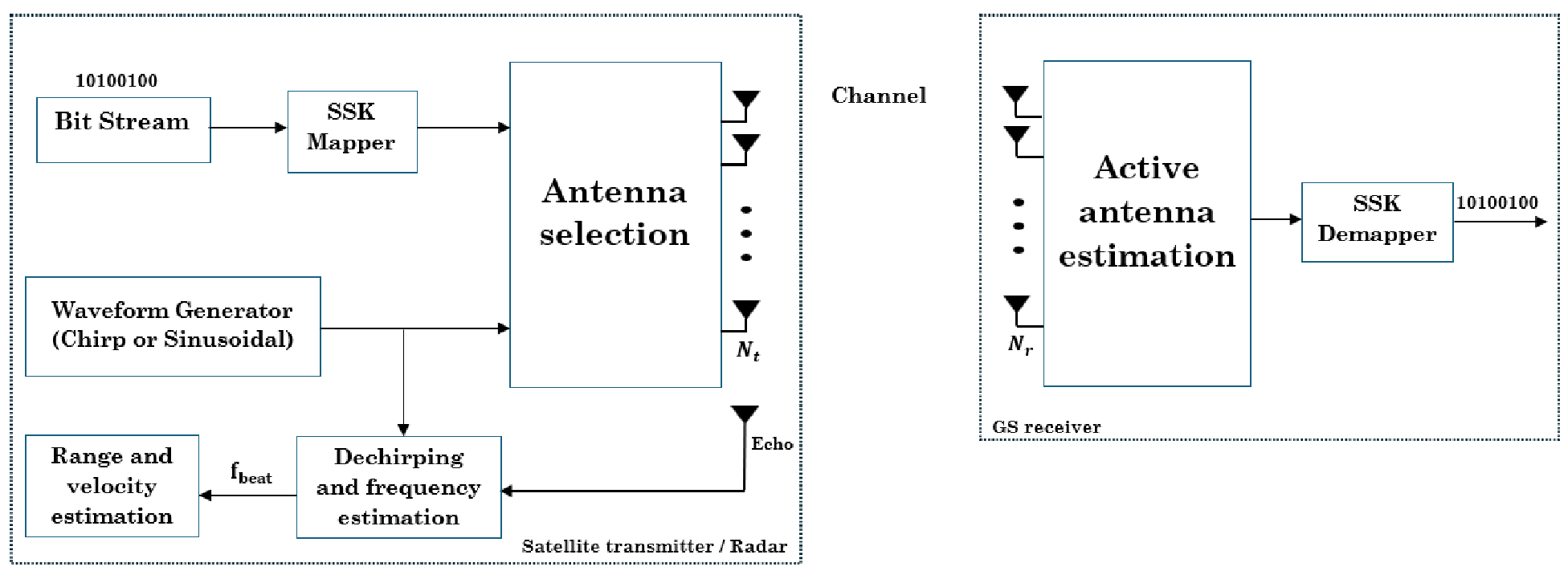} 
    \caption{Block diagram of transmitter and receiver.\vspace{-0.3cm}}
    \label{fig:rx}
\end{figure*}

The LEO satellite transmitter, as illustrated in Fig.~\ref{fig:rx}, is equipped with \(N_t\) transmit  antennas and transmits continuous chirp or sinusoidal 
signal through the channel. 
The transmitter in this system serves dual purposes: data communication and radar sensing. As shown in Fig.~\ref{fig:rx},  it begins with a binary bit stream, which is mapped to specific transmitting antennas using the SSK mapper. The antenna selection ensures efficient spatial transmission of information. Simultaneously, a waveform generator produces either chirp signals, commonly used for radar sensing because of their range and velocity estimation capabilities, or sinusoidal signals for simpler communication purposes. The selected antenna from the array transmits the generated waveform, which not only carries the encoded data but also enables radar functionality by reflecting off objects in the environment for sensing applications. 

\vspace{-0.5cm}
\subsection{Receiver}
Based on our system model depicted  in Fig. \ref{model}, there are two receiving systems: The co-located LEO satellite  which  receives and processes the radar signal reflected by the debris and the GS which detect the transmitted data.
\subsubsection{LEO satellite radar receiver}
The   radar receiver and transmitter are co-located, as illuatrated in Fig. \ref{fig:rx}. The radar receiver system is meticulously engineered to process incoming echo signals reflected by debris, facilitating the extraction of radar sensing metrics. For chirp waveform, the dechirping process, a key component of the system, involves mixing the received signal with a reference chirp. This operation effectively reduces the intermediate frequency bandwidth, facilitating low-speed analog-to-digital conversion and optimizing signal processing.  
This step is critical for ensuring the system's ability to handle high-frequency signals while minimizing hardware demands. To enhance detection accuracy, particularly in low SNR environments, the Root-MUSIC algorithm and fast Fourier transform (FFT) are used to extract the beat frequency (the difference between the reference signal and the received signal frequency). These advanced algorithms enable precise beat frequency estimation,  improving the receiver's capacity to detect and interpret signals with accuracy.

\subsubsection{Ground station (GS) receiver}
The GS receiver, equipped with \(N_r\) receive antennas as shown in Fig. \ref{fig:rx}, is designed to detect the  data transmitted over activated transmit antenna indices. It can receive data on space debris detection from LEO transmitter satellite, providing essential insights for collision avoidance and space traffic management. 
The main task of the receiver is the active antenna estimation, which involves identifying the specific transmit antenna activated at the satellite transmitter using the maximum likelihood (ML) detector. This step is crucial for decoding the data modulated using SSK, as the information is encoded in the spatial domain to antenna indices. Once the active antenna is identified, the SSK demapper reconstructs the original bit stream. The receiver's design with $N_r$ antennas provides accurate detection and decoding of the transmitted data, ensuring reliable communication while complementing the radar sensing capabilities integrated into the system.

\textit{Remark:} Unlike conventional schemes that require complex RF chains and linear amplifiers, SSK allows for a single RF chain and avoids amplitude or phase modulation, reducing the hardware burden. Moreover, the passive antenna switching mechanism aligns well with power and weight constraints aboard small satellites. Chirp waveform generation can be achieved with low-cost voltage-controlled oscillators (VCOs), and FFT-based processing for dechirping is compatible with onboard FPGA or digital signal processing implementations. This makes the proposed ISAC framework not only theoretically effective but also practically realizable within modern satellite payload constraints.
\vspace{-0.3cm}
\section{Communication and Debris Detection Analysis}
\subsection{Analysis of SSK-based Communication Link}
In the following, we discuss and analyze the communication link components between the transmitting LEO satellite and the receiving ground station. 
\subsubsection{Encoding Process}
SSK is a spatial modulation technique that uniquely exploits the spatial domain in Multiple Input Multiple Output (MIMO) systems by encoding information into the indices of active transmit antennas. Unlike traditional modulation schemes that vary signal properties such as amplitude, phase, or frequency, SSK encodes data by selectively activating a single antenna at a time, with the active antenna index representing the transmitted data. This activation is directly linked to the spatial domain without modulating the signal's amplitude or phase, which simplifies the transmitter design by eliminating the need for complex constellations like Phase-Shift Keying  or Quadrature Amplitude Modulation.  
In our $N_t$-antenna SSK system, a binary sequence $\mathbf{b} \in \{0,1\}^{\log_2(N_t)}$ is uniquely mapped to one active antenna index. The transmitted signal vector $\mathbf{v}(t)$ is given by $\mathbf{v}(t) = \mathbf{e}_k x(t)$, where $\mathbf{e}_k$ is the $k$-th column of the identity matrix $\mathbf{I}_{N_t} \in \mathbb{R}^{N_t \times N_t}$, and $x(t)$ is the transmitted waveform (e.g., chirp or sinusoidal). 
\subsubsection{Transmitted Waveforms}
Chirp and sinusoidal waveforms are integrated to our SSK-based systems, serving distinct purposes in encoding data and ensuring robust sensing capabilities. 
Chirp-based waveforms, known for their linear frequency variation over time, can be classified as up-chirp or down-chirp \cite{sumen2022lfmwaveform} . In an up-chirp, the frequency increases linearly and can be represented by
\begin{equation}
    x_{\text{up}}(t) = A \cos\left(2\pi f_0 t + \frac{\mu}{2}t^2\right),
\end{equation}
where $f_0$ is the centre frequency, $A$ is the amplitude, and $\mu = \frac{\Delta F}{T}$ defines the chirp rate based on the bandwidth $\Delta F$ and pulse duration $T$. Conversely, a down-chirp features a linearly decreasing frequency, expressed as
\begin{equation}
    x_{\text{down}}(t) = A \cos\left(2\pi f_0 t - \frac{\mu}{2}t^2\right).
\end{equation}
By alternating between up-chirp and down-chirp signals within a single pulse, these waveforms mitigate range-Doppler coupling effects, enabling precise range and velocity estimation for efficient debris detection.  
This simple yet effective encoding ensures consistent communication performance while complementing the chirp waveforms in ISAC systems. Algorithm \ref{alg:algo1}  describes the function used to generate up-chirp and down-chirp waveforms. 
\begin{algorithm}[t]
\caption{Chirp Generator}\label{alg:algo1}
\begin{algorithmic}[1]
    \REQUIRE $\Delta F$, $T$, $fs$, $bit$
    \ENSURE waveform
    \STATE $t \gets 0:1/fs:T - 1/fs$
    \STATE $x_{\text{up}} \gets \text{chirp}(t, \Delta F/2, T, -\Delta F/2, '\text{linear}')$
       $+ i \text{chirp}(t, \Delta F/2, T, -\Delta F/2, '\text{linear}', 90)$
    \STATE $x_{\text{down}} \gets \text{chirp}(t, -\Delta F/2, T, \Delta F/2, '\text{linear}')$
    $ + i\text{chirp}(t, -\Delta F/2, T, \Delta F/2, '\text{linear}', 90)$
        \STATE $\text{waveform} \gets [x_{\text{up}}, x_{\text{down}}]$
    \RETURN $\text{waveform}$
\end{algorithmic}
\end{algorithm}

To show the robustness of chirped waveforms, we comparer the system performance of using chirped waveforms with that using sinusoidal waveform. For sinusoidal signal, the transmitted waveform is expressed as 
   $ x_{\textrm{sin}}(t) = 
    A \sin(2\pi f t)$.
The activated antennas transmit the same waveform $x(t)$ which can be a chirped or sinusoidal waveform such that
\begin{equation}
    x(t) = 
    \begin{cases} 
   x_{\textrm{chirp}}(t), & \text{for chirped waveform}  \\ 
    x_{\textrm{sin}}(t), & \text{for sinusoidal waveform},
    \end{cases}
\end{equation}
where $x_{\text{chirp}}(t)=x_{\text{up}}(t)x_{\text{ down}}(t)$.
\subsubsection{Channel Model}
The communication link, can be divided into path loss and small-scale fading components \cite{zhai2024golden}. Firstly, the path loss, \(PL_{sg}\), including large-scale effects, is expressed as
\begin{equation}
PL_{sg} = L_b + L_g + L_s,
\end{equation}
where \(L_b\), \(L_g\), and \(L_s\) represent the basic path loss, attenuation due to atmospheric gases, and attenuation from ionospheric or tropospheric scintillation, respectively. The basic path loss in dB is modeled as
\begin{equation}
L_b = \text{FSPL}(d, f_c) + \text{SF} + \text{CL}(\theta_E, f_c),
\end{equation}
where FSPL is the free-space path loss, SF denotes shadow fading with Gaussian distribution as \(\text{SF} \sim \mathcal{N}(0, \sigma_{\text{SF}}^2)\), and CL is the clutter loss. The free-space path loss is calculated as 
\begin{equation}
\text{FSPL}(d, f_c) = 32.45 + 20\log_{10}(f_c) + 20\log_{10}(d),
\end{equation}
where \(d\) is the slant distance in meters, \(f_c\) represents the carrier
frequency in GHz, and \(\theta_E\) being the elevation angle between the
GS and  the transmitting LEO satellite.  The slant distance \(d\) is
determined by the satellite altitude \(h_0\) and \(\theta_E\) such that\\
\begin{equation}
d = \sqrt{R_E^2 \sin^2(\theta_E) + h_0^2 + 2h_0 R_E - R_E \sin(\theta_E)},
\end{equation}
where \(R_E\), the Earth's radius, is approximately 6,371 km \cite{zhang2024roles}.
The attenuation due to atmospheric gases, dependent on
frequency, elevation angle, altitude, and water vapor density,
is given by
\begin{equation}
L_g = \frac{A_{\text{zenith}}(f_c)}{\sin(\theta_E)},
\end{equation}
where \(A_{\text{zenith}}\) is the zenith attenuation for frequencies
between 1 and 1000 GHz.
Scintillation loss \(L_s\) arises from ionospheric or tropospheric
effects, relevant for frequencies below 6 GHz (ionospheric)
and above 6 GHz (tropospheric). Rain and cloud attenuation
are also considered for frequencies above 6 GHz, but are
negligible under the clear-sky assumption used here.

Secondly, the small-scale channel is modeled as a shadowed
Rician fading channel with both line-of-sight (LoS) and non-
LoS (NLoS) components. We define \(\mathbf{H_{sg}} \in \mathbb{C}^{N_r \times N_t}\) as the
channel matrix between the LEO transmitting satellite and the GS.
The shadowed Rician fading channel coefficient between the \(i\)-th
transmit antenna and \(l\)-th receive antenna is
\begin{equation}
h_{sg(i,l)} = \sqrt{\frac{K}{K + 1}} |h_{sg(i,l)}^{\text{LoS}}| + \sqrt{\frac{1}{K + 1}} |h_{sg(i,l)}^{\text{NLoS}}|,
\end{equation}
where \(K\) is the Rician factor. As such the channel between the transmitting LEO satellite and the GS is  \(\mathbf{H}_{sg} = \{h_{sg(i,l)}\}_{i=1,l=1}^{N_t,N_r}\).
The LoS and NLoS components of \( h_{sg(i,l)} \), i.e., \( h_{sg(i,l)}^{\text{LoS}} \) and \( h_{sg(i,l)}^{\text{NLoS}} \), are given as follows
\[
|h_{sg(i,l)}^{\text{LoS}}| \sim \text{Nakagami}(m, \Omega), \quad \angle h_{sg(i,l)}^{\text{LoS}} \sim \text{Unif}[0, 2\pi),
\]
\[
|h_{i,l}^{\text{NLoS}}| \sim \text{Rayleigh}(\sigma_R), \quad \angle h_{i,l}^{\text{NLoS}} \sim \text{Unif}[0, 2\pi),
\]
where \(m\) and \(\Omega\) are the shape and spread parameters for the Nakagami distribution, and \(\sigma_R\) relates to the average magnitude of the NLoS component. 
Thirdly, the Doppler shift \(f_{d,sg}\) in the downlink between the satellite and GS is
\begin{equation}
f_{d,sg} = \left( \frac{v}{c} \right) \left( \frac{R_E}{R_E + h_0} \cos \theta_E \right) f_c,
\end{equation}
where \(c\) is the speed of light and \(v\) is the relative velocity between the LEO transmitting satellite and the GS. For a ditance $d$ between the satellite and GS, we can obtain time delay \(\tau\) as \(\tau = \frac{d}{c}\).
By considering the Doppler shift, the channel response between the \(i\)-th transmit antenna and \(l\)-th receive antenna can be obtained as
\begin{equation}
h_{sg(i,l)} (t) = h_{i,l} e^{-j2\pi f_c \tau t} \delta (\tau - \tau_t),
\end{equation}
where \(\tau_t\) represents the delay at \(t\)-th time slot.
\subsubsection{Detection at GS}

After obtaining the system configurations and channel models, the expression of received signals can be given as
\begin{equation}
\mathbf{y}_{\text{sg}} = \sqrt{E_s PL_{sg}}\mathbf{H}_{sg}\mathbf{v} + \mathbf{n},
\end{equation}
where \(\mathbf{n} \in \mathbb{C}^{N_r \times 1}\) represents the white Gaussian noise vector with \(\mathbf{n} = \{n_l\}_{l=1}^{N_r}\), \(n_l\) is white Gaussian noise with zero mean and variance \(N_0\), and \(E_s\) is the symbol’s energy, expressed as \\
\begin{equation}
    E_s = \frac{1}{N_t} \sum_{k=1}^{N_t} |\mathbf{v}_k|^2.
\end{equation}
Therefore, the instantaneous SNR at the \(l\)-th receive antenna can be obtained as
\begin{equation}
\gamma_{\text{sg}, l} = \frac{\left|\sqrt{E_s PL_{sg}}h_{sg,l}^v + n_l\right|^2}{N_0},
\label{eq:snr}
\end{equation}
where we define \(\mathbf{H}_{sg}\mathbf{v} = \{h_{sg, l}^v\}_{l=1}^{N_r}\). Besides, considering the accuracy in signal detection, an adaptive weighting scheme is used. The received signal $\mathbf{y_{\text{sg}}}(t)$ is weighted based on the received power $P_r$
\begin{equation}
\mathbf{y}_{sg(w)}(t) = \mathbf{W} \cdot \mathbf{y_{\text{sg}}}(t),
\end{equation}
where $\mathbf{W} = \text{diag}(w_1, w_2, \dots, w_{N_r})$ and $w_i = P_r(i)/\max(P_r)$. This weighting emphasizes stronger signals while suppressing noise-dominated components. Then, the ML detector is applied in our SSK schemes, by minimizing the Euclidean distance between the received signal and all possible transmitted signals, which can be expressed as follows
\begin{equation}
\{\hat{\mathbf{v}}\} = \arg\min_{\mathbf{v}} \left\| \mathbf{y}_{\text{sg(w)}} - \sqrt{E_s PL_{sg}}\mathbf{H}_{sg}\mathbf{v} \right\|^2,
\end{equation}
where \(\hat{\mathbf{v}}\)  represents the estimated \(\mathbf{v}\). After the processing of the ML detector, we can easily demodulate and decode the detected results, and obtain the transmitted bit sequences. The ML detection strategy, combined with adaptive weighting, ensures reliable performance even in challenging channel conditions.

By leveraging SSK modulation and demodulation, the system offers high spectral efficiency, reduced power consumption, and robustness against interference and Doppler effects.
\subsubsection{BER Performance} The conditional error probability (EP) between the transmitted ${\mathbf{v}}$  and the detected signal sets $\hat{\mathbf{{v}}}$ can be given as \cite{Raj2024IRS} 
\begin{equation}
\begin{split}
\text{EP}\left(\{v\} \to \{\hat{v}\}\! \mid\! \{H_{sg}\}\right)\! = \!\Pr\Bigg( \!\!\left\| \mathbf{y}_{\text{sg(w)}} - \sqrt{E_s PL_{sg}}\,\mathbf{H}_{sg}\mathbf{v} \right\|^2 \\
> \left\| \mathbf{y}_{\text{sg(w)}} - \sqrt{E_s PL_{sg}}\,\mathbf{H}_{sg}\hat{\mathbf{v}} \right\|^2 \Bigg).
\end{split}
\label{eq:pe}
\end{equation}

After solving (\ref{eq:pe}) as done in \cite{li2021space} , the conditional EP can be obtained as given by
\begin{equation}
\text{EP}_{\{v\} \to \{\hat{v}
\}} = \mathbb{E}\left[ Q\left( \sqrt{\frac{E_s PL_{sg}}{2} \left\| \mathbf{H}_{sg}(\mathbf{v} - \hat{\mathbf{v}}) \right\|^2} \right) \right],
\end{equation}
where $\mathbb{E}[\cdot]$ is the expectation and $Q(\cdot)$ denotes the Gaussian $Q$-function. 
\vspace{-0.3cm}
\subsection{Analysis of Radar and Echo signals}
The  transmitted radar waveform is considered as linear frequency-modulated continuous wave (FMCW). 
The echo signal received by the radar is given by
\begin{equation}
s_\text{echo}(t) = A_r(t) \exp \left(j 2\pi \left(f_\text{beat} t + \phi \right) \right),
\end{equation}
where 
 $A_r(t)$, $f_\text{beat}$, and $\phi$ are the amplitude term,
    beat frequency,
    and phase term, respectively.
The beat frequency for up-chirp and down-chirp signals can be expressed as
\begin{equation}
f_\text{beat,up} = \frac{\Delta F}{T} \cdot \frac{4R}{c} - \frac{2V_r}{\lambda},
\label{eq:fbup}
\end{equation}
\begin{equation}
f_\text{beat,down} = \frac{\Delta F}{T} \cdot \frac{4R}{c} + \frac{2V_r}{\lambda},
\label{eq:fbdown}
\end{equation}
where 
    $R$, $V_r$, $c$, $\lambda$ are the target range,
    target velocity,
 speed of light,
  and wavelength of the signal, respectively.
  Therefore, due to range-Doppler coupling, ambiguity problem arises in velocity and range estimations of moving target \cite{fmcw}. Triangle LFM and V-LFM pulse waveforms created by combining up-chirp and down-chirp signals eliminate this problem as velocity components cancel each other out in the $f_{\text{beat}}$ equation.

  Complex envelope of triangle LFM pulse and V-LFM pulse can be expressed as
\begin{equation}
\tilde{x}_{\text{tri-LFM}}(t) =
\begin{cases} 
\sqrt{\frac{1}{2T}} \, \text{rect}\left(\frac{t}{2T}\right)e^{j\pi \mu t^2}, & -T < t < 0, \\
\sqrt{\frac{1}{2T}} \, \text{rect}\left(\frac{t}{2T}\right)e^{-j\pi \mu t^2}, & 0 < t < T, \\
0, & \text{else.}
\end{cases}
\end{equation}
\begin{equation}
\tilde{x}_{\text{V-LFM}}(t) =
\begin{cases} 
\sqrt{\frac{1}{2T}} \, \text{rect}\left(\frac{t}{2T}\right)e^{-j\pi \mu t^2}, & -T < t < 0, \\
\sqrt{\frac{1}{2T}} \, \text{rect}\left(\frac{t}{2T}\right)e^{j\pi \mu t^2}, & 0 < t < T, \\

0, & \text{else.}
\end{cases}
\end{equation}
where $\tilde{x}_{\text{tri-LFM}}(t)$ and $\tilde{x}_{\text{V-LFM}}(t)$ refer to the complex envelope of the triangle LFM pulse and V-LFM pulse, respectively. Finally, when combined waveforms are used, the range and velocity can be estimated without ambiguity as in  (\ref{eq:range}) and (\ref{eq:velocity}) by adding/subtracting (\ref{eq:fbup}) and (\ref{eq:fbdown}).
\begin{equation}
\hat{R} = \frac{Tc}{8\Delta F} \left(f_{\text{beat,down}} + f_{\text{beat,up}}\right) 
\label{eq:range}
\end{equation}
\begin{equation}
\hat{V_r} = \frac{\lambda}{4} \left(f_{\text{beat,down}} - f_{\text{beat,up}}\right).
\label{eq:velocity}
\end{equation}
Detection accuracy can be evaluated using the Neyman-Pearson criterion with a threshold derived as
\begin{equation}
    \text{Threshold} = P_n \cdot \gamma, \quad \gamma = Q^{-1}(1 - P_{FA}),
\end{equation}
where \(P_{FA}\) is the false alarm probability, and \(P_n\) is the noise power. 
The accuracy for range and velocity estimation is calculated as
\begin{equation}
R_{\text{accuracy}} (\%) = 100 - \frac{|R - \hat{R}|}{R} \times 100 \tag{15}
\end{equation}
\begin{equation}
V_{\text{accuracy}} (\%) = 100 - \frac{|V_r - \hat{V_r}|}{V_r} \times 100.\tag{16}
\end{equation}
The ambiguity function, which characterizes the range-Doppler resolution of the radar waveform, is defined as
\begin{equation}
|\chi(\tau, f_d)|^2 = \left| \int_{-\infty}^\infty x(t) x^*(t - \tau) e^{-j 2 \pi f_d t} \, dt \right|^2,
\end{equation}
where 
 $\tau$ is the time delay that is proportional to the range and
    $f_d$ is the Doppler frequency shift that is proportional to the velocity.

The proposed waveform achieves superior range-Doppler resolution by combining up-chirp and down-chirp signals, mitigating coupling effects. These metrics validate the robustness of the waveform for space debris detection in dynamic LEO environments.
\vspace{-0.4cm}
\section{Numerical Results}
The theoretical and simulation results of the proposed ISAC system are presented and discussed in this section, along with necessary insights and conclusions. 
Monte Carlo simulations are employed for performance evaluation, with all experiments conducted over 
$1 \times 10^6$
  independent channel realizations to ensure statistical reliability. Consistent with conventional schemes, the SNR is represented as $E_b/N_0$, where 
$E_b = |x_k|^2 = 1$ denotes the energy per symbol.
 The major simulation parameters are listed in Table \ref{tab:simulation_parameters}.
We assume a LOS channel between the LEO satellite and the GS. 
\begin{table}[t]
\centering
\caption{Simulation Parameters.}
\resizebox{1\linewidth}{!}{
\begin{tabular}{|l|l|}
\hline

\text{Parameter}                & \text{Value/Description}                                                                 \\ \hline
\text{Sampling Frequency (\(f_s\))} & \(28.8\ \text{MHz}\)                                                                   \\ \hline
\text{Number of Transmit Antennas (\(N_t\))} & \(4\), \(8\), \(16\), \(32\)                                                                 \\ \hline
\text{SNR Interval [dB]}         & \([0, 25]\)                                                                              \\ \hline
\text{Channel Model}            & \begin{tabular}[c]{@{}l@{}}
                                     \text{Rician Channel}: \\
                                     - Rician factor \(K
 = 10\) dB \\ 
                                     \end{tabular}                                                                             \\ \hline
\text{Noise Variance}           & Calculated from SNR as \(\sigma^2 = 10^{-\text{SNR}/10}\)                                 \\ \hline
\text{Bandwidth (\(BW\))}       & \(10\ \text{MHz}\)                                                                       \\ \hline
\text{Carrier Frequency (\(f_c\))} & \(5\ \text{GHz}\), where \(c = 3 \times 10^8\ \text{m/s}\)                            \\ \hline
\text{Satellite Altitude (\(h_0\))} & \(780\ \text{km}\)                                                                    \\ \hline
\text{Elevation Angle (\(\theta_E\))} & \(60^\circ\), i.e., \(d = 884.85\ \text{km}\)                                       \\ \hline
\text{Shadow Fading (\(\sigma_{SF}\))} & \(1\)                                                                             \\ \hline
\text{Zenith Attenuation (\(A_\text{zenith}\))} & \(0.22\)                                                                  \\ \hline
\text{Scintillation Loss (\(L_s\))} & \(0.13\ \text{dB}\) (worst case)                                                      \\ \hline
\text{Rician Distribution Parameter (\(\sigma_R\))} & \(1\)                                                                \\ \hline
\text{Shape Parameter (\(m\))} & \(0.8\)                                                                                   \\ \hline
\text{Spread Parameter (\(\Omega\))} & \(1\)                                                                               \\ \hline
\text{Doppler Velocity (\(v\))} & \(100\ \text{km/s}\)                                                                     \\ \hline
\text{Doppler Angle (\(\alpha\))} & \(30^\circ\)                                                                           \\ \hline
\text{Distance LEO Sat. - GS (\(d\))} & \(884.85\ \text{km}\)                                                                        \\ \hline
\text{Randomized Debris Parameters} & \begin{tabular}[c]{@{}l@{}}
                                         Range \(R \in [500 , 2500]\ \text{km from LEO sat.}\) \\ 

                                         Velocity \(Vr \in [7 000, 8 500]\ \text{m/s}\) \\
                                         \end{tabular}                                                                      \\ \hline
\end{tabular}}
\label{tab:simulation_parameters}

\end{table}

 \begin{figure}[t]
        \centering
\includegraphics[scale=0.58]
{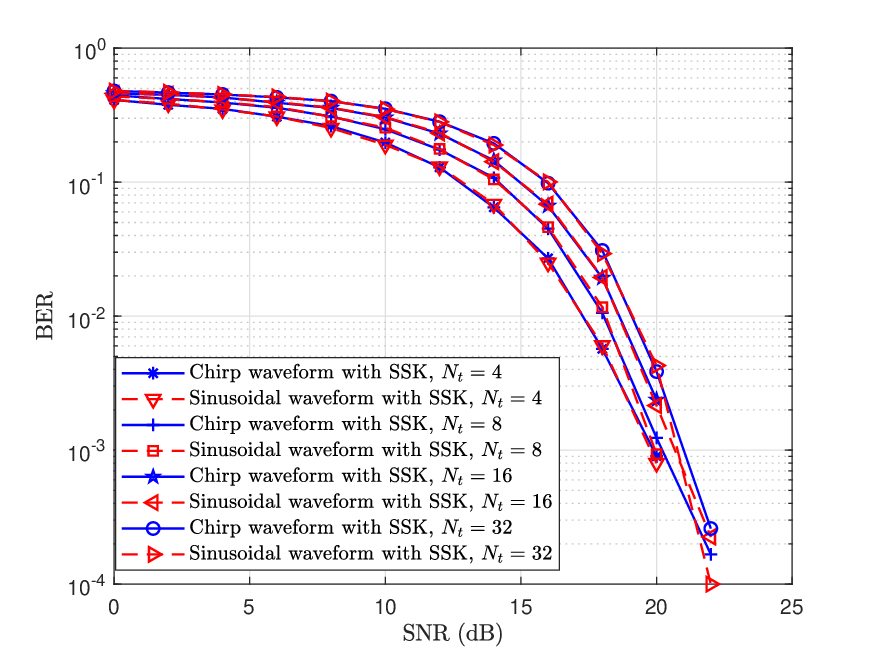}
        \caption{BER performance vs SNR for $N_t$ = 4, 8, 16 and 32, using different types of radar waveforms, i.e, chirp and sinusoidal waveforms.
       \vspace{-0.6cm} }
        \label{fig:g1}
    \end{figure}
The results in Fig. \ref{fig:g1} shows the BER performance of the communication link between the LEO satellite transmitter and the GS receiver.  
 The graph clearly shows that the BER decreases slightly as the number of transmit antennas increases. This is expected since the larger spatial constellation makes it more challenging for the receiver to accurately identify the active antenna, especially in noisy or fading channels. However the spectral efficiency in SSK improves with the number of transmit antennas because more antennas enable encoding a higher number of bits per symbol, $\textrm{log}_2(N_t)$,  without requiring additional bandwidth.
Interestingly, this figure also demonstrates that using SSK, both chirped and sinusoidal waveforms provide approximately equivalent BER performance for a given number of transmit antennas $N_t$.
 This equivalence provides flexibility in selecting the waveform that offers the highest accuracy for sensing or the one that provides simpler radar design.

\begin{figure}[t]
         \centering
         \includegraphics[width=0.45\textwidth]{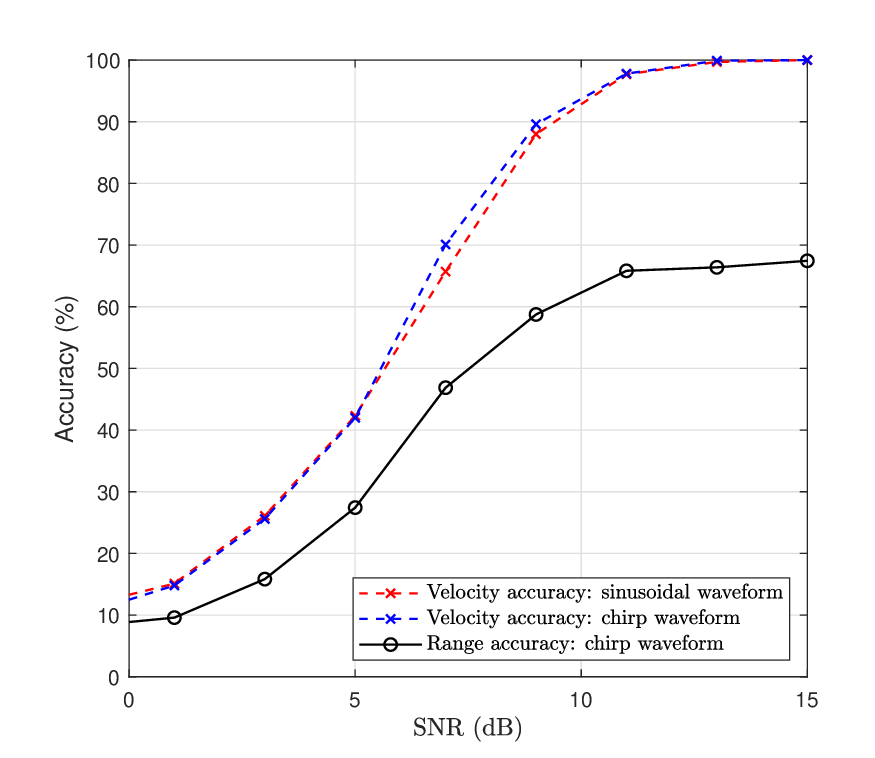}
         \caption{Detection accuracy of  chrip and sinusoidal radar waveforms.
         \vspace{-0.4cm}}
         \label{fig:g3}
     \end{figure}

Fig. \ref{fig:g3} presents the sensing capability of the radar system using sinusoidal and chirp waveforms over varying SNR levels. 
The results show that Both waveforms converge toward near-perfect velocity accuracy $100\%$ as SNR exceeds 10 dB. For range detection, only the chirp waveform is considered, as the sinusoidal waveform is not suitable due to its high ambiguity, which makes accurate range estimation impractical. The range detection accuracy with chirp waveforms improves with increasing SNR but saturates near $70\%$, indicating a performance ceiling even at high SNR. In general, chirp waveforms offer enhanced sensing capabilities, whereas sinusoidal waveforms enable simpler radar design.
 
\vspace{-0.3cm}
\section{Conclusion}

This paper explored the integration of SSK modulation in ISAC systems utilizing different radar waveforms. The findings confirm that SSK supports reliable communication with both chirp and sinusoidal waveforms, that achieve comparable BER performance. However, the waveform choice plays a critical role in sensing performance. While the sinusoidal waveform benefits from a simpler design, its high ambiguity makes it unsuitable for range detection. In contrast, the chirp waveform enables  range sensing and offers a slight improvement in velocity detection accuracy. These results highlight the complementary roles of SSK and waveform design in ISAC systems—where SSK ensures communication efficiency, and waveform selection determines sensing effectiveness. Together, they offer design flexibility for scalable and mission-adaptable ISAC solutions, particularly in applications such as orbital debris detection.
\vspace{-0.3cm}
\section*{Acknowledgment}
The authors gratefully acknowledge the support from the Canadian Space Agency (CSA) for this work. This research would not have been possible without the valuable resources
provided by the CSA, allowing the authors to advance innovation and exploration in the field of space science and technology.

\vspace{-0.4cm}
\footnotesize
\bibliographystyle{IEEEtran}
\bibliography{reference.bib}
\end{document}